# Sulphur Ion Implantations into Condensed CO$_2$: Implications for Europa

D.V. Mifsud[1,2,†], Z. Kaňuchová[3,†], P. Herczku[2], Z. Juhász[2], S.T.S. Kovács[2], G. Lakatos[2,4], K.K. Rahul[2], R. Rácz[2], B. Sulik[2], S. Biri[2], I. Rajta[2], I. Vajda[2], S. Ioppolo[5], R.W. McCullough[6], and N.J. Mason[1,2]

1  *Centre for Astrophysics and Planetary Science, School of Physical Sciences, University of Kent, Canterbury CT2 7NH, United Kingdom*

2  *Institute for Nuclear Research (Atomki), Debrecen H-4026, Hungary*

3  *Astronomical Institute, Slovak Academy of Sciences, Tatranská Lomnica SK-059 60, Slovakia*

4  *Institute of Chemistry, University of Debrecen, Debrecen H-4032, Hungary*

5  *School of Electronic Engineering and Computer Science, Queen Mary University of London, London E1 4NS, United Kingdom*

6  *Department of Physics and Astronomy, School of Mathematics and Physics, Queen's University Belfast, Belfast BT7 1NN, United Kingdom*

† Corresponding authors:    D.V. Mifsud:    dm618@kent.ac.uk
                           Z. Kaňuchová:   zkanuch@ta3.sk

**ORCID Identification Numbers:**

| | |
|---|---|
| D.V. Mifsud | 0000-0002-0379-354X |
| Z. Kaňuchová | 0000-0001-8845-6202 |
| P. Herczku | 0000-0002-1046-1375 |
| Z. Juhász | 0000-0003-3612-0437 |
| S.T.S. Kovács | 0000-0001-5332-3901 |
| G. Lakatos | 0000-0002-3174-602X |
| K.K. Rahul | 0000-0002-5914-7061 |
| R. Rácz | 0000-0003-2938-7483 |
| B. Sulik | 0000-0001-8088-5766 |
| S. Biri | 0000-0002-2609-9729 |
| I. Rajta | 0000-0002-5140-2574 |
| I. Vajda | 0000-0001-7116-9442 |
| S. Ioppolo | 0000-0002-2271-1781 |
| R.W. McCullough | 0000-0002-4361-8201 |
| N.J. Mason | 0000-0002-4468-8324 |

**Abstract**

The ubiquity of sulphur ions within the Jovian magnetosphere has led to suggestions that the implantation of these ions into the surface of Europa may lead to the formation of $SO_2$. However, previous studies on the implantation of sulphur ions into $H_2O$ ice (the dominant species on the Europan surface) have failed to detect $SO_2$ formation. Other studies concerned with similar implantations into $CO_2$ ice, which is also known to exist on Europa, have offered seemingly conflicting results. In this letter, we describe the results of a study on the implantation of 290 keV $S^+$ ions into condensed $CO_2$ at 20 and 70 K. Our results demonstrate that $SO_2$ is observed after implantation at 20 K, but not at the Europa-relevant temperature of 70 K. We conclude that this process is likely not a reasonable mechanism for $SO_2$ formation on Europa, and that other mechanisms should be explored instead.

*Keywords:*   Europa, $SO_2$, sulphur ions, radiation chemistry, astrochemistry, planetary science

**Plain Language Summary**

$SO_2$ ice is known to exist at the surface of one of Jupiter's moons; Europa. However, the method by which this ice forms is still uncertain. Due to the orbit of Europa being within the giant magnetosphere of Jupiter, it has been proposed that sulphur ions within the magnetosphere could implant into the cold surface ices on Europa and subsequently react to form $SO_2$. However, laboratory experiments looking into the implantation of such ions into $H_2O$ ice (the dominant ice on Europa's surface) and $CO_2$ ice have either failed to yield $SO_2$ or have provided inconclusive results. We have therefore performed an experiment in which we have implanted high-energy sulphur ions into $CO_2$ ice at two temperatures. Our results indicate that such implantations are unlikely to be the mechanism by which the $SO_2$ on Europa is formed, and that other chemical processes should be considered instead.

**Highlights / Key Points**

- Sulphur ions were implanted into $CO_2$ ices at 20 and 70 K to simulate Jovian magnetospheric radiation chemistry at the surface of Europa.
- $SO_2$ was observed to be among the radiolytic products at 20 K, but not at the more Europa-relevant temperature of 70 K.
- Alternative explanations for the formation of $SO_2$ on the surface of Europa should be considered.

**1. Introduction**

Sulphur is a ubiquitous element in the cosmos and is an important participant in many biochemical, atmospheric, and geochemical processes (Mifsud *et al.* 2021a). Within our own Solar System, sulphur astrochemistry is perhaps best associated with the Galilean moon system of Jupiter. The innermost of the Galilean moons, Io, is the most volcanically active celestial body in the Solar System; emitting approximately one tonne of sulphurous molecules per second (Thomas *et al.* 2004). Some of these molecules are subsequently stripped from the Ionian exosphere and dissociated and ionised within the Jovian magnetosphere, whereupon the resultant sulphur ions may interact with the surfaces of the other, icy Galilean moons (Cooper *et al.* 2001). The detection of $SO_2$ ice on the trailing hemisphere of Europa by the *International Ultraviolet Explorer* spacecraft (Lane *et al.* 1981) led many researchers to suggest that an influx of magnetospheric sulphur ions could contribute to its formation there as part of a wider radiolytic sulphur cycle (Carlson *et al.* 1999), in which sulphur is chemically transformed to various molecular forms as a result of its interaction with ionising radiation.

Indeed, the distribution of $SO_2$ on the Europan trailing hemisphere is such that it adopts a so-called bulls-eye pattern (Carlson *et al.* 2005, Hendrix *et al.* 2011, Becker *et al.* 2022). This is suggestive of an exogenic sulphur source in which magnetospheric sulphur ions implanting into the surface ices yield $SO_2$ as a primary product. However, laboratory experiments considering the implantation of such ions into $H_2O$ ices, which represent the dominant icy species at the surface of Europa, have thus far failed to detect any $SO_2$ among the radiolytic products. Instead, $H_2SO_4$ and its hydrates are efficiently formed (Strazzulla *et al.* 2007, Strazzulla *et al.* 2009, Ding *et al.* 2013). Computational simulations have also been unsuccessful in demonstrating a reasonable reaction pathway by which sulphur ion implantations yield $SO_2$ as a primary product (Anders and Urbassek 2019a, Anders and Urbassek 2019b).

Sulphur ion implantations into other oxygen-bearing ices known to be present on the surface of Europa have also been considered more recently. Lv *et al.* (2014) performed experiments that demonstrated that such implantations into pure CO and $CO_2$ ices at 15 K did indeed yield $SO_2$, and proposed that a geologically reasonable time-scale of $2\times10^4$ years is sufficient to produce the abundance of $SO_2$ observed on Europa. However, this estimation relied on the assumption that sulphur ion implantation experiments carried out at 15 K are representative of analogous processes occurring on the surface of Europa, which is characterised by significantly higher temperatures (Ashkenazy 2019). Follow-up studies by Boduch *et al.* (2016) did not detect $SO_2$ after the implantation of sulphur ions into pure $CO_2$ ice at 16 K, clearly contrasting with the previous results of Lv *et al.* (2014). It should be noted, however, that Boduch *et al.* (2016) made use of ultraviolet absorption spectroscopy as their product detection method, and it is therefore possible that the formation of any $SO_2$ was masked by the stronger absorptions of sulphur oxyanions which were detected after implantation. Nonetheless, it is apparent that the question of a possible exogenic sulphur source for $SO_2$ on the surface of Europa remains an open one.

We have therefore made an experimental attempt to address the possibility of $SO_2$ formation as a result of sulphur ion implantation into a carbon oxide ice. In this letter, we present the results of high-fluence ($>10^{16}$ ions cm$^{-2}$) implantations of 290 keV $S^+$ ions into pure $CO_2$ ices at 20 and 70 K. By performing our experiment at two different temperatures (one that is similar to previous experimental work conducted in this field, and one that is relevant to some of the colder Europan surface ices at the mid- to high-latitudes), we have been able to determine the influence of temperature on the radiation chemistry taking place within our ices and thus extend the results described by previous studies. Our results and their implications are discussed in light of the radiolytic chemistry present on the Europan surface.

## 2. Materials and Methods

Our experiments were carried out using the Ice Chamber for Astrophysics-Astrochemistry (ICA) located at the Institute for Nuclear Research (Atomki) in Debrecen, Hungary. The technical details of this set-up have been extensively described in previous publications (Herczku *et al.* 2021, Mifsud *et al.* 2021b), and so only a short overview is provided here. The ICA is an ultra-high vacuum compatible steel chamber containing a gold-coated oxygen-free copper sample holder at its centre. This sample holder is able to host a maximum of four ZnSe substrates on which astrophysical ice analogues may be prepared. The substrates may be cooled to 20 K using a closed-cycle helium cryostat, although an operational temperature range of 20-300 K is available. The base pressure of the chamber is maintained in the region of a few $10^{-9}$ mbar by the combined use of a dry rough vacuum pump and a turbomolecular pump. The chamber is also connected to a 2 MV Tandetron accelerator (Rajta *et al.* 2018, Biri *et al.* 2021) which allows ion beams to be delivered to the prepared astrophysical ice analogues at a nominal incidence angle of 36° to the normal. A simplified schematic diagram of the ICA is given in Figure 1.

$CO_2$ ices were prepared on the ZnSe substrates at 20 and 70 K *via* background deposition of the gas (Linde, 99.995% purity), which was dosed into the main chamber through an all-metal needle valve at a chamber pressure of a few $10^{-6}$ mbar. Ice growth and processing were monitored *in situ* using Fourier-transform mid-infrared (FTIR) transmission absorption spectroscopy over a spectral range of 4000-650

cm$^{-1}$ using a nominal resolution of 1 cm$^{-1}$. Acquired FTIR spectra of the CO$_2$ ices allowed for quantitative measurements of their molecular column densities $N$ (molecules cm$^{-2}$) to be made by using Eq. 1:

$$N = \frac{\int \tau(\nu)\, d\nu}{A_\nu} = \frac{P \ln(10)}{A_\nu}$$

(Eq. 1)

where $\tau(\nu)$ is the optical depth of the molecular ice at a given wavenumber (cm$^{-1}$), $A_\nu$ is the integrated band strength constant for the infrared absorption band over which Eq. 1 is integrated (cm molecule$^{-1}$), and $P$ is the measured area of this absorption band (cm$^{-1}$). The molecular column density of a deposited ice may be related to its thickness $d$ (μm) through Eq. 2:

$$d = 10^4 \frac{NZ}{N_A \rho}$$

(Eq. 2)

where $Z$ and $\rho$ are respectively the molar mass (44 g mol$^{-1}$) and density ($\rho_{20\,K} = 0.98$ g cm$^{-3}$; $\rho_{70\,K} = 1.48$ g cm$^{-3}$) of the deposited CO$_2$ ice (Satorre et al. 2008), and $N_A$ is the Avogadro number (6.02×10$^{23}$ molecules mol$^{-1}$).

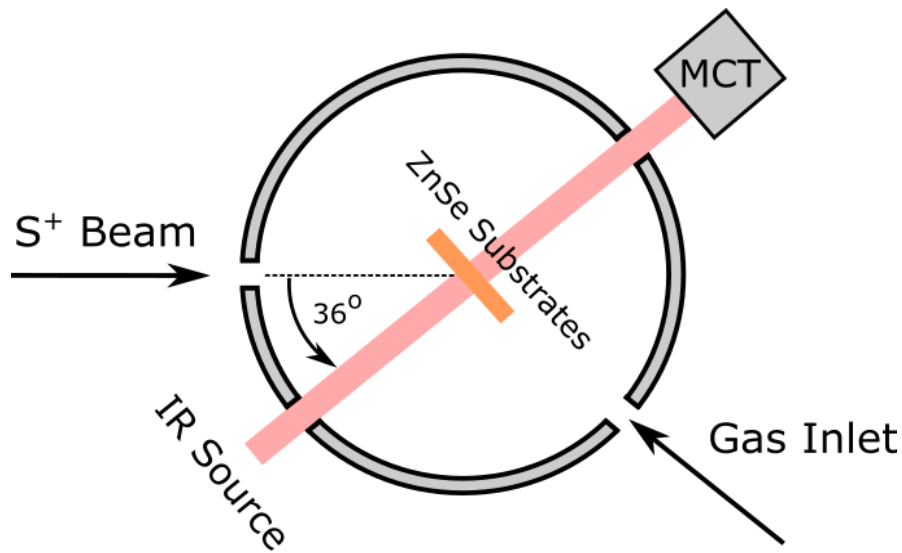

**Figure 1:** Simplified top-view schematic diagram of the ICA set-up. Implantations of S$^+$ ions were performed using the arrangement shown here, with ions impacting the ice targets at angles of 36° to the normal. The FTIR spectroscopic beam, detected by an external MCT detector, remained orthogonal to the substrates and ices. Other attachments to the side of the chamber (e.g., Faraday cups, pressure gauges, etc.) are omitted for clarity.

The most prominent infrared absorption feature of condensed CO$_2$ is the asymmetric stretching mode at around 2343 cm$^{-1}$ (Isokoski et al. 2013, Mifsud et al. 2022a). However, this band saturates fairly quickly during ice deposition; indeed, saturation occurs before the ice reaches a suitable thickness for ion implantation to be performed. Therefore, we have made use of the less intense absorption band related to the asymmetric stretching mode of the $^{13}$CO$_2$ isotopologue at around 2283 cm$^{-1}$ to quantify the column densities and thicknesses of the deposited ices. To do this, a FTIR spectrum of the depositing ice was acquired just prior to the saturation of the $^{12}$CO$_2$ asymmetric stretching mode and the column density ratio of the $^{12}$CO$_2$ and $^{13}$CO$_2$ isotopologues was measured using Eq. 1 assuming $A_\nu$ for the asymmetric stretching modes of these isotopologues to be $7.6\times10^{-17}$ and $7.8\times10^{-17}$ cm molecule$^{-1}$, respectively (Gerakines *et al.* 1995). Continued deposition then resulted in the saturation of the $^{12}CO_2$ asymmetric stretching mode, after which the measured $^{13}CO_2$ column densities were used in combination with the experimental isotopologue abundance ratio to determine the total molecular column densities of the deposited ices as well as their thicknesses (Eq. 2). A similar approach had been previously adopted by Lv *et al.* (2014).

Pure $CO_2$ ices were deposited to thicknesses of about 3 μm. This thickness is greater than the penetration depths of the projectile 290 keV $S^+$ ions supplied by the Tandetron accelerator as calculated using the SRIM programme (0.8-1.1 μm; Ziegler *et al.* 2010), thus ensuring implantation of the ions into the ices. Once an ice was deposited to this desired thickness, a pre-irradiation FTIR spectrum was acquired after which the ice was exposed to the ion beam with further spectra acquired at pre-defined fluence steps until a total fluence of >$10^{16}$ ions cm$^{-2}$ was implanted. During implantation, however, it was noted that sputtering resulted in a gradual thinning of the ice. To compensate for this sputtering, a simultaneous deposition-irradiation method was used. In this method, after initial deposition of the ice to a thickness of 3 μm, the ices were irradiated by the $S^+$ ion beam with concurrent background deposition of more $CO_2$ occurring at a chamber pressure of about $10^{-5}$ mbar. Both irradiation and deposition were halted during FTIR spectral acquisition. We do not anticipate the use of this simultaneous deposition-irradiation method to impact the chemical evolution of our ices any differently to what would be expected under standard ion implantation conditions, and have provided evidence supporting this claim in the appendix section of this letter.

## 3. Results and Discussion

In this study, the implantation of 290 keV $S^+$ ions into $CO_2$ ices at 20 and 70 K was investigated with the aim of determining whether such a mechanism could account for the formation of $SO_2$ on the cold surface of Europa. $CO_2$ ice is a minor component of the Europan surface (Hansen and McCord 2008), having an estimated abundance of 0.036% by number relative to $H_2O$ (Hand *et al.* 2007). As such, it is likely to interact to at least some extent with incident magnetospheric sulphur ions. The fluxes of keV-MeV $S^{n+}$ ions vary between $2\times10^6$ and $10^8$ ions cm$^{-2}$ s$^{-1}$, depending on the location on the surface of Europa (Dalton *et al.* 2013). Therefore, a fluence of $10^{16}$ ions cm$^{-2}$ as was supplied in this study could be expected to be delivered to the Europan surface within 160 years.

The implantation of 290 keV $S^+$ ions into $CO_2$ ice at 20 K resulted in the appearance of several new absorption features in the FTIR spectrum (Figure 2). In particular, two new bands were observed at 1336 and 1150 cm$^{-1}$ which we have respectively attributed to the asymmetric and symmetric stretching modes of $SO_2$ (Sandford *et al.* 1991, Gomis and Strazzulla 2008). These absorption bands are, however, fairly small and are located in a region of the spectrum where several other, more intense absorption bands are also present. In light of this, we have performed two control experiments so as to be confident in our detection of $SO_2$. Firstly, we compared the wavenumber peak positions of these bands with those of $SO_2$ in an unirradiated $CO_2$:$SO_2$ (6:1) ice mixture (Figure 2), and found a very good agreement (within 1 cm$^{-1}$) with the suspected $SO_2$ band positions in the irradiated ice. Secondly, we implanted a similar fluence of 300 keV $He^+$ ions into pure $CO_2$ ice at 20 K under similar experimental conditions. After such implantations, absorption bands due to the sulphur chemistry initiated by the implantation of $S^+$ ions would be absent while all other bands produced as a result of the energetic processing of the ice would still be present. This outcome was indeed observed, with the bands at 1336 and 1150 cm$^{-1}$ not being observed at any point during the implantation of the $He^+$ ions.

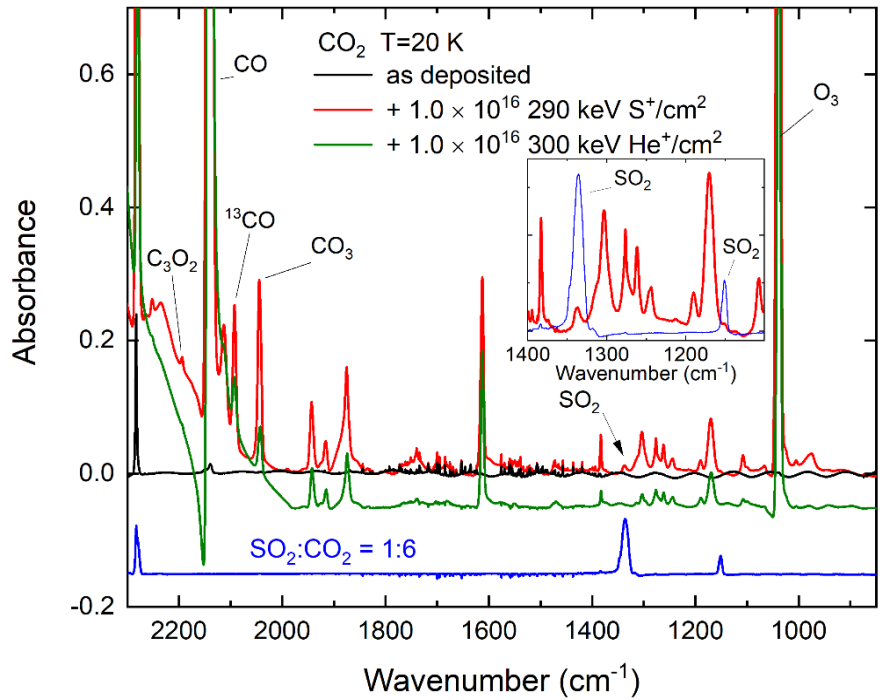

**Figure 2:** FTIR spectra of condensed $CO_2$ before (black trace) and after (red trace) the implantation of 290 keV $S^+$ ions at 20 K. Also shown are the FTIR spectra acquired during control experiments, including an unirradiated $CO_2$:$SO_2$ (6:1) ice mixture at 20 K (blue trace) and a $CO_2$ ice after the implantation of 300 keV $He^+$ ions at 20 K (green trace). The stopping powers of the $S^+$ and $He^+$ ions in the $CO_2$ ice are respectively 4.33 and 9.74 eV Å$^{-1}$, while the radiation doses supplied after delivering $10^{16}$ ions cm$^{-2}$ are 325 and 730 eV per $CO_2$ molecule. Note that the spectra of the irradiated ices are difference spectra yielded after the subtraction of the 'as deposited' spectra.

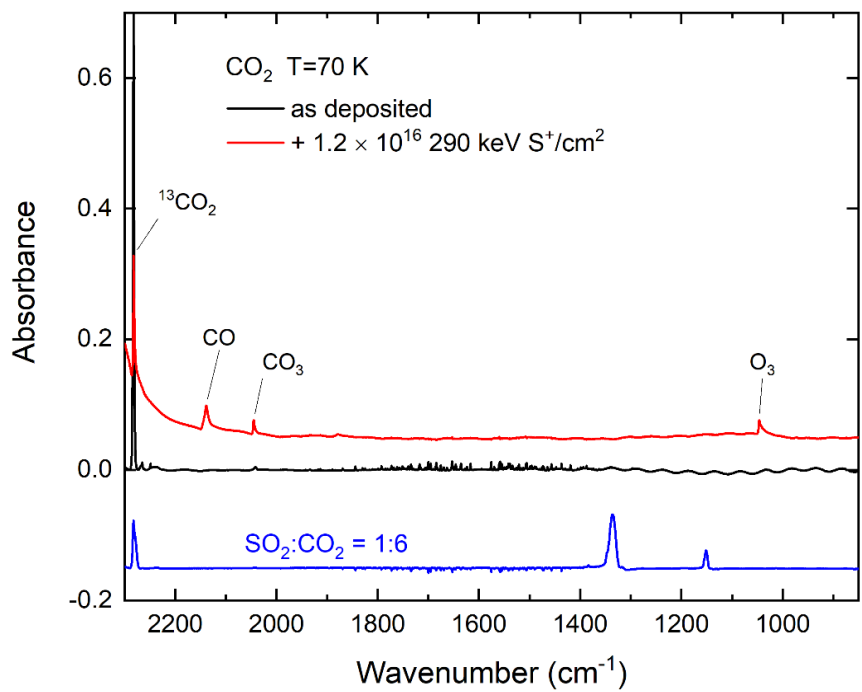

**Figure 3:** FTIR spectra of condensed $CO_2$ before (black trace) and after (red trace) the implantation of 290 keV $S^+$ ions at 70 K. Also shown is the FTIR spectrum of an unirradiated $CO_2$:$SO_2$ (6:1) ice mixture at 20 K (blue trace). Note that the spectrum of the irradiated ice is a difference spectrum yielded after the subtraction of the 'as deposited' spectrum.

The mechanistic chemistry leading to the formation of $SO_2$ is presumed to proceed after the neutralisation of the implanted sulphur ion. As the ion traverses the ice it dissipates energy into its surroundings resulting in the dissociation of $CO_2$ to CO and an oxygen atom (Pilling *et al.* 2022). The neutralised sulphur ion may then react with one such oxygen atom to produce SO (Tevault and Smardzewski 1978); alternatively, SO may result from the abstraction of an oxygen atom from a surviving $CO_2$ molecule by the neutralised sulphur ion (Froese and Goddard 1993). SO is a very reactive species, and so will rapidly oxidise to yield $SO_2$ (Rolfes *et al.* 1965, Herron and Huie 1980).

Interestingly, when the implantation of 290 keV $S^+$ ions was performed at 70 K, no $SO_2$ was detected in the FTIR spectrum (Figure 3). Indeed, this spectrum shows significantly fewer absorption bands than does its 20 K counterpart. One striking difference is the size of the absorption band due to the asymmetric stretching mode of $O_3$ located at about 1041 cm$^{-1}$ (Chaabouni *et al.* 2000). In the 20 K spectrum (Figure 2), this band is very intense and is indicative of the efficient formation of $O_3$ in the irradiated $CO_2$ ice (Mifsud *et al.* 2022a, Mifsud *et al.* 2022b). In the 70 K spectrum (Figure 3), however, this band is significantly smaller, suggesting an inefficient $O_3$ formation pathway. It is this observation that gives us an insight into why the formation of $SO_2$ as a result of $S^+$ ion implantation into condensed $CO_2$ is favourable at 20 K, but not at 70 K.

The energetic processing of $CO_2$ ice by ions, electrons, and ultraviolet photons is known to yield several oxygen-bearing products, including $O_3$ (Sivaraman *et al.* 2013, Martín-Doménech *et al.* 2015, Mifsud *et al.* 2022a). The formation of this radiolytic product is, however, dependent upon the prior synthesis of $O_2$ within the ice which then furnishes $O_3$ upon the barrierless addition of a supra-thermal oxygen atom. In our 20 K experiment, the $O_2$ is stable within the ice as the experimental temperature is lower than its sublimation temperature, and so $O_3$ formation may occur efficiently. Indeed, $O_3$ is one of the major products of this implantation process (Figure 2). In the 70 K experiment, however, it is evident that $O_2$ sublimation from the ice is fairly efficient (Jones *et al.* 2014), thus effectively depleting the ice of its oxygen content. The result of this is that there are fewer oxygen atoms available within the bulk ice that may react with the implanted sulphur to yield $SO_2$, thus explaining the absence of this latter species at 70 K. Previous studies have suggested that otherwise volatile molecules such as $O_2$ may be stabilised at temperatures beyond their sublimation point *via* their encapsulation within clathrate-like structures (Hand *et al.* 2006). We speculate that, in our experiments, any such $CO_2$-based clathrate-like structures may not have been sufficiently stable to retain $O_2$ due to the fact that $CO_2$ is itself a volatile species. The presence of a radiolytically-derived tenuous $O_2$ exosphere on Europa (Milillo *et al.* 2016) is consistent with this interpretation.

Our results build upon and extend the previous findings of Lv *et al.* (2014) and Boduch *et al.* (2016). In their study, Lv *et al.* (2014) implanted multiply charged sulphur ions into condensed CO and $CO_2$ at 15 K, and recorded the formation of $SO_2$ in each ice. Based on their reported $SO_2$ formation efficiency of 0.38 molecules ion$^{-1}$ for the implantation of 90 keV $S^{9+}$ ions into $CO_2$ ice, Lv *et al.* (2014) suggested that the observed abundances of $SO_2$ on Europa could be formed within a geologically reasonable time-scale of 2×10$^4$ years. This result was not reproduced in the later study by Boduch *et al.* (2016), who did not observe any $SO_2$ in the ultraviolet absorption spectra of $CO_2$ ice into which 144 keV $S^{9+}$ ions had been implanted at 16 K. This non-detection was ascribed to one of two reasons: either the accumulated column density of $SO_2$ formed as a result of ion implantation was below the spectroscopic detection limits of their instrument, or else the absorption bands attributable to $SO_2$ were masked by the intense absorptions of sulphur oxyanions such as $SO_3^-$.

The results of this present study suggest that, although the formation of $SO_2$ as a result of sulphur ion implantation into $CO_2$ ice is indeed possible at low temperatures (15-20 K), this is not true for higher temperatures more representative of the surface of Europa, such as the 70 K temperature considered here. This is actually a somewhat unexpected result, as other ion implantation processes in which the implanted ion is incorporated into an oxygen-bearing product molecule have been shown to be unaffected by changes in the experimental temperature (Lv *et al.* 2012, Ding *et al.* 2013). We conclude, therefore, that sulphur ion implantation into $CO_2$ ices at the surface of Europa is not an efficient mechanism by which the $SO_2$ observed on the surface may form. This is consistent with the lack of association in the spatial distributions of $SO_2$ and $CO_2$ ices on Europa (Hansen and McCord 2008). Other formation mechanisms must therefore be considered instead. The correlation in the distribution of $SO_2$ and $H_2SO_4$ hydrates on the surface of Europa, as well as their bulls-eye patterns, hints at a related synthetic chemistry. Indeed, an early study by Hochanadel *et al.* (1955) demonstrated that the irradiation of concentrated $H_2SO_4$ solution yielded $SO_2$, and a similar result was more recently observed by Loeffler *et al.* (2011) for condensed $H_2SO_4$ and its hydrates at low temperatures. Therefore, it is reasonable to suggest that the implantation of sulphur ions into the Europan $H_2O$ surface ices yields $H_2SO_4$ hydrates (Strazzulla *et al.* 2007, Strazzulla *et al.* 2009, Ding *et al.* 2013), which are then dissociated to yield $SO_2$ upon further irradiation. It is also possible that alternative formation mechanisms, such as the radiolytic decomposition of sulphate minerals (McCord *et al.* 2001) and the implantation of cold (sub-keV) magnetospheric sulphur ions into the Europan surface ice (Becker *et al.* 2022), may further contribute to the presence of $SO_2$.

## 4. Conclusions

In this letter, we present the results of a study considering the implantation of 290 keV $S^+$ ions into condensed $CO_2$ at 20 and 70 K as a possible mechanism by which the $SO_2$ ice on the surface of Europa may form. We have found that, although $SO_2$ is observed amongst the radiolytic products after implantation of the ions into $CO_2$ ice at 20 K, it is not observed after implantation at 70 K; a temperature more relevant to the Europan surface. We have attributed this to the fact that, at this higher temperature, any $O_2$ formed as a result of the combination of radiolytically-derived oxygen atoms will efficiently sublime from the bulk ice, thus depleting it of the oxygen necessary for $SO_2$ synthesis. Based on the results of previous studies that have demonstrated the efficient synthesis of $H_2SO_4$ and its hydrates after sulphur ion implantation into $H_2O$ ice, as well as the observed spatial correlation between these hydrates and $SO_2$ on the surface of Europa, we suggest that a possible major source of $SO_2$ may be the radiolytic dissociation of $H_2SO_4$ hydrates, themselves formed as a result of the implantation of sulphur ions into the Europan surface.

## 5. Appendix

The above experiment made use of a simultaneous deposition-irradiation system in which pure $CO_2$ ices were first deposited to a thickness of about 3 μm, after which the ices were irradiated with a 290 keV $S^+$ ion beam with concurrent background deposition of further $CO_2$ ice at a chamber pressure of $10^{-5}$ mbar. It is possible that the interaction of the ion beam with gas-phase $CO_2$ molecules may result in the formation of fragment species that may then deposit onto the growing ice and potentially influence any solid-phase radiation chemistry occurring there. However, under the experimental conditions described above, we consider the incorporation of such fragments into the depositing ice to be a negligible process and, in this appendix, we provide calculations to support this statement.

The expected number $n_e$ of $CO_2$ molecules undergoing fragmentation as a result of their collision with $S^+$ ions (molecules ion$^{-1}$) is given as:

$$n_e = \sigma L \rho_{mol}$$

(Eq. A1)

where $\sigma$ is the fragmentation cross-section of the molecule (cm$^2$ ion$^{-1}$), $L$ is the pathlength of the ion in the chamber before it collides with the ice layer (cm), and $\rho_{mol}$ is the molecular gas density (molecules

cm$^{-3}$). A value for this latter term may be calculated by first considering the Ideal Gas Law, which gives the ratio of the amount of gas present $n$ (mol) to its volume $V$ (cm$^3$) to be:

$$\frac{n}{V} = \frac{P}{RT} = 4 \times 10^{-7} \text{ mol m}^{-3}$$

(Eq. A2)

where $P$ is the gas pressure (Pa; $10^{-5}$ mbar = $10^{-3}$ Pa), $R$ is the molar gas constant (8.314 J K$^{-1}$ mol$^{-1}$) and $T$ is the temperature of the gas which we have taken to be 298 K (i.e., room temperature). Converting this value to the molecular gas density $\rho_{\text{mol}}$ (molecules cm$^{-3}$) may be achieved by multiplying by the Avogadro number (6.02×10$^{23}$ molecules mol$^{-1}$):

$$\rho_{\text{mol}} = N_A \frac{n}{V} = 2.4 \times 10^{17} \text{ molecules m}^{-3} = 2.4 \times 10^{11} \text{ molecules cm}^{-3}$$

(Eq. A3)

Assuming that $\sigma = 10^{-15}$ cm$^2$ ion$^{-1}$ (Mejía *et al.* 2015) and taking $L = 20$ cm (known from the geometry of the chamber) and substituting into Eq. A1, a value of 4.8×10$^{-3}$ molecules ion$^{-1}$ may be calculated for $n_e$. Consider now the rate of ion delivery to the ICA chamber, $r_{\text{ion}}$ (ions s$^{-1}$): this may be determined by taking the ratio of the nominal S$^+$ beam current ($I = 100$ nA) to the Coulombic charge of the projectile S$^+$ ions ($q = 1.602 \times 10^{-19}$ C). This yields a value of 6.2×10$^{11}$ ions s$^{-1}$ for $r_{\text{ion}}$. The rate of molecular dissociation to yield fragments, $r_{\text{diss}}$ (molecules s$^{-1}$) is therefore given as:

$$r_{\text{diss}} = n_e r_{\text{ion}} = 3 \times 10^9 \text{ molecules s}^{-1}$$

(Eq. A4)

The pumping speed of the turbomolecular pump maintaining the vacuum in the ICA is 400 L s$^{-1}$ (equivalent to 0.4 m$^3$ s$^{-1}$). Such a value is valid for N$_2$; however, the pumping speed of CO$_2$ gas $S$ is estimated to be 90% of this value and is thus 0.36 m$^3$ s$^{-1}$. By once again making use of the Ideal Gas Law, the partial pressure of those molecules undergoing dissociation due to interaction with the S$^+$ beam, $P_p$, may be given as:

$$P_p = \frac{r_{\text{diss}} RT}{S N_A} = 3.4 \times 10^{-11} \text{ Pa} = 3.4 \times 10^{-13} \text{ mbar}$$

(Eq. A5)

Following Dalton's Law of Partial Pressures, the ratio of the total number of gas-phase CO$_2$ molecules in the chamber at any one time to the number of dissociated CO$_2$ molecules is in excess of 3×10$^7$. As such, it is highly unlikely that the fragments formed by molecular interaction with the S$^+$ ion beam could contribute in any significant way to the radiation chemistry and physics occurring within the irradiated bulk ice (for which the column density is in excess of 4×10$^{18}$ molecules cm$^{-2}$), even if all the fragments formed were to deposit and be incorporated into the ice (a scenario which in of itself is unlikely, since at least some of the fragments should be pumped out of the chamber by the turbomolecular and dry rough vacuum pumps before they can condense). Therefore, we conclude that the simultaneous deposition-irradiation method used in this study was appropriate.

**Acknowledgements**

The authors acknowledge support from the Europlanet 2024 RI which has received funding from the European Union's Horizon 2020 Research Innovation Programme under grant agreement No. 871149. The main components of the ICA set-up were purchased using funds from the Royal Society obtained through grants UF130409, RGF/EA/180306, and URF/R/191018. Further developments of the installation were supported in part by the Eötvös Loránd Research Network through grants ELKH IF-2/2019 and ELKH IF-5/2020. This work has also received support from the European Union and the State of Hungary; co-financed by the European Regional Development Fund through grant No. GINOP-2.3.3-15-2016-00005. Support has also been received from the Research, Development, and Innovation Fund of Hungary through grant No. K128621. DVM is the recipient of a University of Kent Vice-Chancellor's Research Scholarship. The research of ZK is supported by the Slovak Grant Agency for Science (grant No. 2/0059/22) and the Slovak Research and Development Agency (contract No. APVV-19-0072). ZJ is grateful for the support of the Hungarian Academy of Sciences through the János Bolyai Research Scholarship. SI acknowledges the Royal Society for financial support.

**Author Contributions**

The experiment was designed by DVM and carried out by DVM, PH, ZJ, STSK, and BS. Additional experimental support was received from GL and KKR. Data analysis was performed by ZK. The calculations presented in the appendix section of this letter were performed by ZJ and BS. All authors contributed to results interpretation. The manuscript was written by DVM, and all authors were responsible for its improvement.

**Declaration of Interests Statement**

The authors declare that they have no known interests (financial or otherwise) that may have influenced the interpretation of the results presented in this study.

**Open Research**

The raw spectroscopic data collected during experimentation is available on the Mendeley Data online repository: https://doi.org/10.17632/zr2mdbk5dt.1 (Mifsud *et al.* 2022c). Data analysis was performed using the OriginPro 2018 software: https://www.originlab.com/.